# Universality of the Mott-Ioffe-Regel limit in metals

N. E. Hussey[1], K. Takenaka[2] and H. Takagi[2,3]
[1]*H. H. Wills Physics Laboratory, University of Bristol, Tyndall Avenue, Bristol, BS8 1TL, U. K.*
[2]*RIKEN (Institute of Physical and Chemical Research), Wako-shi, Saitama 351-0198, Japan.*
[3]*Department of Advanced Materials Science, Graduate School of Frontier Sciences, University of Tokyo, Kashiwa-no-ha 5-1-5, Kashiwa-shi, Chiba 277-8651, Japan.*

The absence of resistivity saturation in many strongly correlated metals, including the high-temperature superconductors, is critically examined from the viewpoint of optical conductivity measurements. Coherent quasiparticle conductivity, in the form of a Drude peak centred at zero frequency, is found to disappear as the mean free path (at $\omega = 0$) becomes comparable to the interatomic spacing. This basic loss of coherence at the so-called Mott-Ioffe-Regel (MIR) limit suggests that the universality of the MIR criterion is preserved even in the presence of strong electron correlations. We argue that the shedding of spectral weight at low frequencies, induced by strong correlation effects, is the primary origin of the extended positive slope of the resistivity to high temperatures observed in *all* so-called 'bad metals'. Moreover, in common with those metals which exhibit resistivity saturation at high temperatures, the scattering rate itself, as extracted from optical spectra, saturates at a value consistent with the MIR limit. We consider possible implications that this ceiling in the scattering rate may have for our understanding of transport within a wide variety of bad metals and suggest a better method for analysing their optical response.

**1. INTRODUCTION**

Our most basic definition of a metal is a material whose resistivity $\rho$ increases with temperature $T$. Within the quasiparticle concept, this is equivalent to saying that the mean-free-path $\ell$, i.e. the average distance a quasiparticle travels between collisions, becomes progressively shorter as the number of thermally-induced scattering events grows. In ordinary metals, phonons are the dominant scatterer at high temperatures and accordingly, $\rho(T)$ is found to increase linearly with $T$ above some typical phonon energy. In considering the problem of low mobilities in certain semiconductors, Ioffe and Regel (1960) realized that a metallic state cannot survive this reduction in $\ell$ indefinitely. They argued that $\ell$ can never become shorter than the interatomic spacing $a$, since at that point the concept of carrier velocity is lost and all coherent quasiparticle motion vanishes. Similar arguments were later expressed by Mott (1972) and the notion of a minimum metallic conductivity compatible with a minimum mean free path $\ell_{\rm min} = a$ [Gurvitch 1981] became known as the Mott-Ioffe-Regel (MIR) limit. The precise numerical definition of the limit is not generally agreed upon; criteria ranging from $k_F\ell_{\rm min} \sim 1$ through $\ell_{\rm min} \sim a$ to $k_F\ell_{\rm min} \sim 2\pi$ have all been employed (here $k_F$ is the Fermi wave vector), but for the purposes of this paper we assume it to be $\ell_{\rm min} \sim a$.

Perhaps the most striking manifestation of the MIR criterion (i.e. $\ell \geq a$) is the phenomenon of resistivity saturation. In certain highly-resistive elements, alloys and intermetallic compounds, $\rho(T)$ deviates from linearity at high $T$, its $T$-dependence becoming progressively weaker as $\rho(T)$ approaches a constant value in the region 100 – 200 μΩcm consistent with $\ell_{\mathrm{min}} \sim a$ [Mooij 1973, Fisk and Webb 1976]. In many compounds, this approach to saturation is found to follow a simple parallel-resistor formula (PRF) [Wiesmann *et al.* 1977],

$$\frac{1}{\rho(T)} = \frac{1}{\rho_{\mathrm{ideal}}(T)} + \frac{1}{\rho_{\mathrm{sat}}} \tag{1}$$

Here, $\rho_{\mathrm{ideal}}$ is the ideal resistivity that would be expected in the absence of saturation, e.g. from Boltzmann transport theory, and $\rho_{\mathrm{sat}}$ the maximum resistivity corresponding to $\ell_{\mathrm{min}} = a$.

Table I contains a selection of elements and compounds for which equation (1) has been applied (in each case successfully) to describe the functional form of $\rho(T)$ in its approach to saturation. The second and third columns give respectively the corresponding values of $\rho_{\mathrm{sat}}$ and $\ell_{\mathrm{min}}$ (in units of $a$). The list is by no means exhaustive and while there have been numerous attempts to explain the origin of the PRF (for an excellent critique of the exisitng theoretical literature, see Allen (1980), Gurvitch (1983) and Gunnarsson *et al*. (2003)), its ability to describe the approach to the MIR limit does not appear in doubt. Indeed, in his review of resistivity saturation, Allen (1980) remarked that both the PRF and resistivity saturation (at the MIR limit) should be universal properties of *all* metals, irrespective of the strength of coupling to the scattering entities. In good metals of course, $\ell >> a$ at all experimentally accessible temperatures (right up to the melting point $T_m$). Hence, $\rho_{\mathrm{ideal}}$ is never more than a fraction of $\rho_{\mathrm{sat}}$ and the full resistivity curve can be adequately described by a conventional Boltzmann-type approach. Even in good metals however there are instances where the *electron* temperature has been elevated above $T_m$, e.g. in Al using pulsed lasers [Milchberg 1988], and $\rho(T)$ once again is found to saturate at a value consistent with $\ell_{\mathrm{min}} \sim a$.

While at the time Allen's conjecture appeared genuine, the physical significance of the PRF with its apparent second conduction channel remained controversial. The original arguments of Mott, Ioffe and Regel had been developed for elastic scattering in disordered materials. It was not immediately obvious why a high scattering rate caused by strong *inelastic* scattering at high temperatures would have a similar effect. Moreover, around the same time as the experiment of Milchberg *et al*. (1988) on Al, a new form of metallicity was emerging that seemed to challenge all previous reasoning. In the then newly discovered high temperature superconductors, $\rho(T)$ was found to grow approximately linearly with $T$ up to $T_m$ ~ 1000 K *without any sign of saturation*. In the process, $\rho(T)$ attained values between 1 and 10 mΩcm, more than one order of magnitude higher than $\rho_{\mathrm{sat}}$ of typical metals, corresponding to $\ell << a$ [Gurvitch and Fiory 1987]. Since this clearly invalidated conventional Boltzmann theory, the absence of resistivity saturation was swiftly acknowledged as a possible signature of novel non-Fermi liquid (non-FL) behaviour.

Such non-saturating $\rho(T)$ has since been registered in a wide range of strongly correlated systems, both inorganic and organic. Collectively, they have come to be referred to as 'bad metals' [Emery and Kivelson 1995]. Their anomalous behaviour has reinvigorated the debate on resistivity saturation and new theoretical ideas have begun to emerge. Most recently, the inclusion of strong correlation effects, both within dynamical mean-field theory (DMFT) [Merino and McKenzie 2000] and the *t-J* model [Calandra and Gunnarsson 2003], have been demonstrated to yield resistivities beyond the MIR limit. In the process, the importance of extending the analysis into the frequency (optical) domain, that covers a much wider range of (photon) energies up to and beyond the bandwidth $W >> k_BT_m$, has been recognised [Rozenberg *et al*. 1995, Merino and McKenzie 2000, Calandra and Gunnarsson 2003].

In this paper we explore further the issue of resistivity saturation from the perspective of optical conductivity data. We argue that whilst the DMFT and *t-J* models can explain several aspects of the evolution of $\sigma(T,\omega)$ spectra in bad metals, the key feature of their optical response is in fact a suppression of low-frequency spectral weight that sets in well before $\rho(T)$ shows any sign of saturation [Kostic *et al.* 1998, Takenaka *et al*. 2002a, b]. An empirical relation between the loss of coherence and a critical value of $\sigma(T,0)$ is established [Takenaka *et al*. 2002a, b], making the MIR criterion relevant to the transport dynamics of all metals, even those with strong electron correlations. As we shall show, the term 'bad metal' turns out to be something of a misnomer. In the search for a possible explanation for this loss of coherence, we consider the various energy scales and scattering processes which distinguish those metals that exhibit resistivity saturation and those that do not. Finally, we examine extending the PRF-style analysis into the optical domain, in both saturating and non-saturating metals, and discuss what implications saturation in the ($\omega$-dependent) scattering rate $\Gamma(T,\omega)$ has on our understanding of the intrinsic nature of $\Gamma(T,\omega)$ in high-$T_c$ cuprates.

The paper is organised as follows: In section 2, we broaden our discussion of resistivity saturation to encompass the heavy-fermion (HF) metals for which $\sigma(T,\omega)$ data are available over a broad range of temperatures and frequencies. The observation of *conductivity* saturation (at the MIR limit) in HF systems is particularly significant, as will become apparent in section 3 where we investigate the 'violation' of the MIR criterion in other strongly correlated metals. In section 4, we analyse the relative importance of different factors that can lead to bad metallic behaviour and extend the PRF analysis into the realm of finite frequencies. Our concluding remarks are presented in section 5.

## 2. HEAVY FERMION METALS

The optical conductivity of a metal can be decomposed into a collection of Lorentz harmonic oscillators plus a Drude peak to low $\omega$. The former contributions arise from absorption at finite frequency, e.g. due to vibrational infrared active modes (phonons) and/or to electronic interband transitions. The Drude term describes the coherent quasiparticle contribution to the electrodynamic response. The peak is centred at $\omega = 0$ but extends out to frequencies of order the carrier containing bandwidth $W$ from which point interband transitions dominate $\sigma(\omega)$. In a conventional metal, the Drude peak broadens with increasing $T$, with a width at half-maximum equal to the mean scattering rate $\Gamma$, but crucially, persists for all $T < T_m$. The issue of what happens to the Drude term as one approaches the MIR limit, i.e. in the regime where $\Gamma$ becomes of order $W$, has received little attention, largely due to the fact that in typical saturating metals, the saturation regime appears well above the temperature limit for conventional optical spectrometers. Indeed, complementary optical measurements have yet to be performed on any of those systems listed in table 1. There is one family of compounds however, namely the HF metals, where $\rho(T)$ does saturate well below room temperature and therefore whose optical response can be studied in the saturation limit. Being strongly correlated systems themselves, the HF metals also form a unique bridge between metals that show resistivity saturation and the bad metals, to be discussed in section 3, that do not.

At high $T$, HF metals behave as a weakly interacting ensemble of localized *f*-electrons and mobile *d*-electrons of moderate bandwidth and effective mass. As $T$ is lowered below a certain coherence temperature $T_{coh}$, the *f*-electron spins become strongly coupled to the conduction electrons and to one another via exchange spin-flip. This coupling leads to the development of an extremely sharp resonance in the density of states at the Fermi level $\varepsilon_F$ and strong upward renormalization of the conduction-electron effective mass $m^*$ by as much as three orders of magnitude [Fisk *et al*. 1988].

Table 2 lists various parameters extracted from $\rho(T)$ and $\sigma(T,\omega)$ measurements on three representative HF compounds, the 5*f* compounds $UPt_3$ and $UPd_2Al_3$ and 4*f* $CeAl_3$. The electrical resistivities of these compounds show the generic behaviour of HF metals; a surprisingly low residual resistivity $\rho_0$, a large $T^2$ coefficient at low $T$ and a saturation that sets in well below 300 K with $\rho_{sat}$ values (60 - 200 μΩcm) comparable to those seen in other saturating metals (see table 1). In some cases, the resistivity also shows a broad maximum around $T = T_{coh}$, presumably due to competing $T$-dependencies in $n$, $m^*$ and $\Gamma$ around $T_{coh}$.

Optical conductivity measurements carried out on these systems reveal intriguing signatures of conductivity saturation in the frequency domain. In $UPt_3$ and $UPd_2Al_3$ for example, $\sigma(T,\omega)$ at 300K is characterized by a flat, essentially featureless spectrum that extends all the way up to $W$ ~ 0.3 eV [Sulewski *et al*. 1988]. The magnitude of the conductivity there is ~ 5000 and 6000 $(\Omega\text{cm})^{-1}$ respectively, in good agreement with the observed dc $\rho_{sat}$. Whilst in $CeAl_3$, optical conductivity data have only been taken up to $T_{coh}$ = 10K [Geibel *et al*. 1991], $\rho(T)$ at this temperature is approaching already the saturation limit and the 10K spectra show many features qualitatively similar to those seen in the 300K spectra of $UPt_3$ and $UPd_2Al_3$.

In analyzing such spectra, it has become increasingly common practice to employ the so-called *generalized* or *extended* Drude analysis which introduces a *$\omega$-dependent* scattering rate $\Gamma(\omega)$ [Allen and Mikkelsen 1977]. For HF metals, $\Gamma(\omega) \sim \omega^2$ at low $T$ and low $\omega$, mirroring the $T^2$ form of the resistivity. At high $\omega$, $\Gamma(\omega)$ is found to saturate at a value $\Gamma_{max}$ where $W/2 < \Gamma_{max} < W$ (see table 2). As $T$ increases, $\Gamma(\omega)$ merges into $\Gamma_{max}$ though $\Gamma_{max}$ itself does not change appreciably with temperature. This latter observation is particularly significant as it implies that $\Gamma_{max}$ represents some fundamental ceiling in the scattering rate of such saturating metals. The flat spectrum at high $T$ can thus be viewed as direct evidence for $\Gamma$ becoming comparable (though not exceeding) $W$.

The essence of the optical response of HF metals was captured in a seminal work by Millis and Lee (1987) in which they considered a band of nearly free electrons hybridizing with a highly correlated band of *f*-electrons within a lattice Anderson Hamiltonian in the Kondo limit (see Degiorgi (1999) for a review). At low $T$ and $\omega$, where impurity scattering dominates,

$$\sigma(\omega) = \frac{ne^2}{m_b}\frac{\tau}{1+(m^*/m_b)^2\omega^2\tau^2} = \frac{ne^2}{m^*}\frac{\tau^*}{1+(\omega\tau^*)^2} \qquad (2)$$

where $m_b$ is the bare optical mass and $1/\tau^* = 1/\tau(m_b/m^*)$ is the renormalized scattering rate in the highly-correlated coherent state that gives rise to the narrow Drude response.

At larger $\omega$, the coupling of itinerant (heavy) electrons to the spin-fluctuation spectrum leads first to a strong $\omega^2$-dependence in the scattering rate and ultimately to a frequency-independent (i.e. saturated) conductivity of the form

$$\sigma(\omega) \sim \frac{ne^2}{m_b}\frac{1}{N(m^*/m_b)\varepsilon_F^*} \sim \frac{ne^2}{m_b W} \qquad \text{(at low } T\text{).} \qquad (3)$$

Here $N$ is the orbital degeneracy of the $f$ state and $\varepsilon_F{}^*$ the renormalized Fermi energy of order the Kondo temperature $T_K$. Note that this form for the conductivity corresponds to a flat spectrum of width $W$, consistent with a scattering rate $\Gamma_{max} \sim W$ (i.e. the MIR limit). This is also precisely the dc resistivity at high temperatures.

Whilst the strongly $T$-dependent quasiparticle renormalization in HF metals restricts the applicability of the PRF to these systems, they nevertheless represent a particularly extreme and revealing example of resistivity (conductivity) saturation whose significance has previously been overlooked. Firstly, both the dc and optical data confirm the MIR limit as being the relevant threshold in HF systems. Secondly, as $T$ increases, the Drude tail simply broadens with $T$ until eventually it evolves into a 'plateau' of width $\omega \sim W$. Once saturation has been reached, $\sigma(T,\omega)$ *shows no further qualitative change in behaviour with increasing T* (provided $T < W$). In essence this is the ultimate manifestation of conductivity saturation. What is more, the height and width of the $\sigma(\omega)$ plateau appear consistent *with a complete preservation of the intraband spectral weight*. As will be shown in the following section, this is the key difference between the optical response of HF systems and what is seen in bad or non-saturating metals.

## 3. NON-SATURATING METALS

### 3.1 Resistivity

As stated in the Introduction, a growing number of metallic compounds are found to show resistivity behaviour that is markedly distinct from that seen in the A15s or the HF metals. They include the high-$T_c$ cuprates, the manganites and vanadates, the ruthenate family and the organic salts. They often lie in close proximity to a charge-ordered or magnetically-ordered ground state and as a result, are associated with many of the more interesting phenomena in current solid state research, including unconventional superconductivity, colossal magnetoresistance and non-FL behaviour. Table 3 lists a sample of these non-saturating metals, together with estimates of the maximum resistivity $\rho_{met}$ at which 'metallic' behaviour has been observed. (For the sake of our discussion, we define a non-saturating metal here as one in which no qualitative change in $\rho(T)$ is observed as it crosses the threshold corresponding to the MIR criterion. Thus, whilst saturation of sorts (i.e. a change in slope) is observed in both the high-$T_c$ cuprates [Takagi *et al*. 1992] and the manganites [Takenaka *et al*. 2002b], it occurs at such high values of resistivity, that we consider it to be of different origin to the type of resistivity saturation described in sections 1 and 2). A quick inspection of tables 1 and 3 reveals that $\rho_{met}$ can be more than one order of magnitude larger than $\rho_{sat}$ of typical saturating metals.

Whilst the high-$T$ resistive behaviour of these bad metals may appear compatible with our original definition of a metal, the sub-Angstrom mean-free-paths one extracts from $\rho(T)$ render physically meaningless any conventional picture of Bloch waves propagating coherently through a periodic potential commensurate with the underlying crystal lattice. Moreover, given the rather smooth $T$-dependence of $\rho(T)$ across the MIR threshold, it has been claimed that the absence of resistivity saturation in bad metals must be a direct manifestation of their non-FL ground state at 0K [Emery and Kivelson 1995]. At first sight, this point of view appears rather compelling. Significantly however, quantum oscillatory phenomena *of the FL form* have now been observed both in organic salts [Wosnitza *et al*. 1996] and two members of the ruthenate family, $SrRuO_3$ [Mackenzie *et al*. 1998] and $Sr_2RuO_4$ [Mackenzie *et al*. 1996], proving that a well-defined FL ground state is formed in (at least some of) these systems at low $T$. Evidence for a coherent, three-dimensional FL ground state has also been reported recently for highly doped cuprates [Proust *et al*. 2003, Hussey *et al*. 2003, Nakamae *et al*. 2003]. Hence, the automatic association of a non-saturating $\rho(T)$ with a non-FL *ground* state is inappropriate and some other physical explanation for the high-$T$ resistivity is required.

### 3.2 Optical conductivity

Optical conductivity data extending to temperatures up to and beyond 300K have by now been collected on most of the compounds listed in table 3 and a generic and rather striking behaviour seems to be emerging. This is illustrated in figure 1 where $\sigma(\omega)$ data for underdoped $La_{1.9}Sr_{0.1}CuO_4$ are reproduced (without the phonon peaks) from Takenaka *et al*. (2003). At low $T$, $\sigma(\omega)$ is dominated by a large Drude-like peak characterized by an anomalous frequency dependence ~ $\omega^{-\gamma}$, ($0.5 \leq \gamma \leq 1$). As $T$ increases, the low-frequency feature initially broadens and $\sigma(0)$ drops, thereby signalling an increase in the quasiparticle scattering rate plus in some cases, a redistribution of the spectral weight.

At a critical temperature $T_{crit}$ (~ 400K for $La_{1.9}Sr_{0.1}CuO_4$) a subtle but fundamental change in the charge dynamics is heralded by the flattening out of $\sigma(\omega)$ in the low frequency limit at a corresponding value $\sigma_{crit}(0)$. Above $T_{crit}$, the plateau in $\sigma(\omega)$ evolves into a dip in the far-infrared limit, the low-frequency spectral weight being transferred to energies $\omega > W$ (~ 1 eV) and the optical sum-rule is only fulfilled at much higher energies of 2 - 3 eV [Takenaka *et al*. 2003]. Note that whilst this low-$\omega$ response is evolving, the spectra at high $\omega$ (viz. $\omega > W/3$) remain largely $T$-independent. The same overall behaviour has now been observed in $La_{2-x}Sr_xCuO_4$, $Bi_2Sr_2CuO_6$ [Tsvetkov *et al*. 1997], $SrRuO_3$ [Kostic *et al*. 1998], $La_{1-x}Sr_xMnO_3$ [Takenaka *et al*. 2002b], $NiS_{1-x}Se_x$ (Miyasaka *et al*., unpublished data) and the organic compounds $\kappa$-$(ET)_2Cu[N(CN)_2]Br$ [Eldridge *et al*. 1991] and $\beta$''-$(ET)_2SF_5CH_2CF_2SO_3$ [Dong *et al*. 1999]. In some of the organics, the zero-frequency Drude peak is never actually realized since $T_{crit}$ turns out to be lower than the minimum temperature at which experiments have been performed [Dong *et al*. 1999]. In all instances however, the position of the dip, created through this transfer of low-frequency weight, is found to shift to higher $\omega$ as $T$ continues to rise and further weight is redistributed from $\omega$ ~ 0 to $\omega > W$. This behaviour is in marked contrast to that seen in HF metals where spectral weight remains below some cut-off of order $W$ at all temperatures. It is this loss of low-frequency weight above $T_{crit}$ therefore that is driving the dc resistivity beyond the MIR limit in non-saturating metals.

For the crystal shown in figure 1 $\sigma_{crit}(0)$ ~ 1000 $(\Omega cm)^{-1}$. This value lies between the two limits $\ell_{min} = a$ ($\sigma_{min}$ = 1500 $(\Omega cm)^{-1}$) and $k_F\ell = 1$ ($\sigma_{min}$ = 600 $(\Omega cm)^{-1}$) for the 2D Fermi surface predicted for $La_{1.9}Sr_{0.1}CuO_4$. As the doping level is decreased, $T_{crit}$ decreases but significantly, $\sigma_{crit}(0)$ remains constant to within experimental resolution (see table 3) [Takenaka *et al*. 2003]. This striking result suggests that the onset of spectral weight loss is inextricably linked with the approach to the MIR limit. As shown in the inset to figure 1, these marked changes in $\sigma(T,\omega)$ are not reflected in $\rho(T)$; $\rho(T)$ simply maintains its 'metallic' climb across either coherent/incoherent boundary, implying that the crossover from Boltzmann to diffusive transport beyond the MIR limit is a gradual one [Nikolic and Allen 2001, Calandra and Gunnarsson 2002]. Note that this is not always the case however; in $V_2O_3$ [Rozenberg *et al*. 1995], $NiS_{1-x}Se_x$ [Miyasaka *et al*., unpublished data), the manganites [Takenaka *et al*. 2002b] and some of the organics [Limelette *et al*. 2003], $\rho(T)$ is found to increase sharply beyond the MIR limit once the zero-frequency Drude peak is lost.

### 3.3 Comparison with Theory

There are three key features of this generic behaviour that need to be taken into account in any successful theory of bad metallic conductivity; (i) the development of the low-frequency dip and its extension to higher frequencies as $T$ increases, (ii) the relatively $T$-independent $\sigma(\omega)$ spectra at high frequencies and (iii) the values of $T_{crit}$ and $\sigma_{crit}(0)$. In DMFT [Rozenberg *et al*. 1995, Merino and McKenzie 2000, Limelette *et al*. 2003], it is the strong $T$-dependence of the spectral density that drives the system from coherent (FL-like) to incoherent behaviour above a certain crossover temperature $T_0$. This crossover is marked by complete suppression of the Drude peak at $\omega = 0$, a shift of this spectral weight out to higher frequencies, a change in the $T$-dependence of $\rho(T)$ around $T_0$ and ultimately resistivity saturation at values far beyond the MIR limit. Significantly however, the disappearance of the Drude peak at $T \sim T_0$ in DMFT calculations is correlated with the onset of resistivity saturation, in contrast with experiment. Indeed, comparison of figures 3 and 9 in Merino and McKenzie (2000) confirms that the Drude peak is expected to survive even for resistivities much larger than the MIR limit.

Gunnarsson and Calandra (2002, 2003) have considered a scenario based on the *t-J* model to account for the high resistivities found in underdoped cuprates. In their approach, the average kinetic energy of charge carriers is reduced due to the strong Coulomb repulsion on the Cu-O sublattice by a factor $x(1-x)$, where $x$ is the (small) percentage of doped holes in the $CuO_2$ planes. $\sigma(\omega)$ is assumed to evolve at high $T$ into a flat, incoherent spectrum (rather like the one seen in HF metals) of width $W$ and a height $\sigma_{min}$ (<< $\sigma_{min}$(MIR)) that is consistent with the resistivity ceiling for each particular doping. As $T$ is lowered, a Drude tail emerges from this constant background, leading to an increase in $\sigma(0)$ and a concomitant drop in $\rho(T)$. As with the DMFT, a metallic $\rho(T)$ is synonymous with a finite Drude peak at $\omega = 0$.

Table 3 lists values of $\sigma_{crit}(0)$ for those compounds where it has been measured, along with estimates of $\sigma_{MIR}$, the conductivity corresponding to $\ell_{min} = a$. (In cases where it has yet to be measured, an asterix marks the predicted value.) Given the experimental uncertainties, the perceived correlation between $\sigma_{crit}$ and $\sigma_{MIR}$ ought not to be taken too literally at this stage (though further quantitative comparisons are strongly encouraged.) What is clear however is the striking difference between $\sigma_{crit}(0)$ and $\sigma_{met}(0) = 1/\rho_{met}$. The important point here is that *a fundamental crossover in the charge dynamics of bad metals occurs well before any saturation threshold*; $\sigma(0)$ continues to drop with increasing *T due to the gradual removal of spectral weight at low frequencies*; the positive slope of $\rho(T)$ in bad metals beyond the MIR limit is thus deceptive and should in no way be taken as a continuation of the standard metallic state that is observed at lower temperatures.

## 4. DISCUSSION

### 4.1 Possible Origins of Bad Metallic Behaviour

Having established the phenomenology, we now turn to consider which parameters distinguish bad metals from saturating metals, without being specific to any one system, and to explore the various possible origins for bad metallic behaviour itself. As discussed in the previous section, the observation of quantum oscillations in both organic salts [Wosnitza *et al*. 1996] and the ruthenates [Mackenzie 1996, 1998] makes the association of bad metallic behaviour with a non-FL ground state insupportable. Whilst bad metals are believed to be characterized by narrow conduction bands ($W$ < 1eV), this is always not the case. In the manganites, for example, the reflectivity edge can extend out to 2 eV [Takenaka *et al*. 2002b]. Conversely, both the A15s and the HF metals have relatively narrow (high-temperature) bandwidths of order 0.5 eV. The development of the low-frequency dip in $\sigma(T,\omega)$ at high $T$ also rules out both thermal expansion (thought to be significant in the organics and fullerenes, for example, but less so in transition metal oxides) and lattice anharmonicity as explanations for the ‘excess resistivity’ in bad metals. The lack of resistivity saturation in the cubic perovskite $SrRuO_3$ [Allen *et al*. 1996] and the spinel $LiV_2O_4$ [Urano *et al*. 2000] further excludes low dimensionality as a determining factor. Moreover, since resistivity escalation is observed both in single-band metals, like the cuprates and organics, *and* in multi-sheeted $Sr_2RuO_4$, band degeneracy, or the absence of interband scattering, does not seem to be playing a significant role. Finally, the small Fermi surface arguments used to explain the elevated $\rho_{sat}$ values found in some underdoped cuprates [Takagi *et al*. 1992] are not applicable to $Sr_2RuO_4$ for example.

Indeed, the only truly *universal* feature of bad metals is their proximity to a Mott insulating state and the corresponding dominance of electron-electron correlations in their transport behaviour. This is manifest in their large $T^2$ resistivities (at low $T$) extending in many cases over several decades in temperature [Crusellas *et al.* 1991, Kobayashi *et al.* 1993, Hussey *et al*. 1998, Mackenzie *et al*. 1998, Miyasaka *et al.* 2000, Okuda *et al.* 2000, Urano *et al.* 2000, McBrien *et al.* 2002]. (Whilst a large $T^2$ resistivity is only clearly noticeable in electron-doped cuprates [Crusellas *et al.* 1991] or hole-doped cuprates with doping levels beyond the superconducting dome [Nakamae *et al.* 2003], manifestations of strong electron-electron scattering in optimally doped cuprates are seen, for example, in the dramatic rise of the infra-red and thermal conductivities below $T_c$ [Yu *et al.* 1992, Bonn *et al.* 1993].) In saturating metals by contrast, resistivity is dominated by scattering off (bosonic) excitations that are *independent* of the conduction electrons. In the HF metals of course, this is only true at high $T$ (i.e. in the saturation limit) where the dominant scattering is believed to be from spin fluctuations off localized *f*-electron moments [Fisk *et al.* 1988]. In elemental metals and the A15s, the bosons in question are phonons whilst in the Fe alloy $\gamma$-$Fe_{80-x}Ni_xCr_{20}$, electron-magnon scattering appears to be dominant [Nath and Majumbar 1996].

It was noted [Kostic *et al.* 1998, Takenaka *et al.* 2002a, b] that the behaviour of $\sigma(\omega)$ in bad metals at high $T$ is intriguingly similar to that seen in disordered materials at low $T$, where a similar finite-energy edge develops and shifts to higher frequencies *with decreasing temperature* as $\sigma(0)$ is driven down by Anderson localization [Gold *et al*. 1982]. Of course, Anderson localization is not expected to occur at high $T$ since Anderson localization is a quantum interference phenomenon that is disturbed by strong inelastic scattering. A more appropriate description of this suppression of low-frequency spectral weight is in terms of the destruction of the quasiparticle itself, as championed by DMFT [Georges *et al*. 1996, Merino and McKenzie 2000,]. The correlation between $\sigma_{crit}(0)$ and $\sigma_{MIR}$ highlighted here however implies that it is the level of scattering, rather than temperature, that primarily heralds the collapse of the quasiparticle pole, though the two energy scales may of course be closely correlated. In DMFT for example, the coherence temperature $T_0$ is associated with the same renormalized Fermi temperature $\varepsilon_F$* that appears in the enhanced prefactor of the $T^2$ resistivity [Georges *et al*. 2003]. As we approach the MIR limit with increasing temperature, strong correlation effects drive the system across the coherent/incoherent boundary. The presence of a large on-site Coulomb repulsion $U$ pulls spectral weight from low energies up to energies beyond the bandwidth, i.e. into the upper Hubbard band, and as a result the dc conductivity passes below the MIR limit. In this sense, the lack of resistivity saturation ought now to be viewed as direct evidence for proximity of the system to a Mott transition.

**4.2 Application of PRF Analysis to Optical Data**

The observation of some fundamental ceiling in $\Gamma(\omega)$ in the optical response of HF metals (at the MIR limit) suggests that it is the *saturation of the scattering rate* $\Gamma$ at $\Gamma_{max}$ (= $1/\tau_{min}$) which is responsible for saturation of the dc resistivity, rather than any thermally-induced changes in $n$ and/or the plasma frequency $\Omega_p$. Support for this approach comes from disorder studies of resistivity saturation, particularly those where additional scattering has been induced through irradiation [Gurvitch *et al*. 1978, Caton and Viswanathan 1982], a process that has negligible effect upon $n$. For each level of irradiation, the PRF form to $\rho(T)$ is found to hold. More importantly both $\rho_{\text{sat}}$ *and* the $T$-dependent part of $\rho_{\text{ideal}}(T)$ remain unchanged to within experimental accuracy, even for $\rho_0$ values close to $\rho_{\text{sat}}$.

Intriguingly, the same phenomenon is also observed in bad metals. Extended Drude analysis carried out on various cuprates up to high frequencies for example have revealed a $\Gamma(T,\omega)$ that saturates too at a value 3000 - 4000 cm$^{-1}$ ~ $W/2$ comparable with the MIR limit [Jehl *et al.* 1992, Quijada *et al.* 1999, van der Marel *et al.* 2003]. This observation is particularly intriguing in cuprates where $\rho(T)$ is seen to attain values significantly higher than one would expect for such maximal levels of scattering. It also hints that the notion of a minimum scattering time (maximum scattering rate) is indeed universal.

These empirical observations, coupled with the success of the PRF in describing $\rho(T)$ in saturating metals, lead us to consider whether a PRF-style analysis can also be applied to the optical response. Before proceeding however, we reiterate that any analysis based on two conducting channels is inappropriate and the term 'parallel-resistor formula' is actually rather misleading. In what follows therefore, we adopt the more relevant term 'scattering rate saturation' or SRS for short. We start by incorporating the MIR criterion into the electron scattering probability distribution, as proposed originally by Gurvitch (1981). Rather than thinking in terms of a single *mean* scattering time $\tau$ for all quasiparticles, Gurvitch considered the probability d$P$ of an individual quasiparticle to scatter in a time interval ($t$, $t$ + d$t$) as given by

$$\mathrm{d}P = \frac{1}{\tau} e^{-t/\tau} \mathrm{d}t \tag{4}$$

Then, by invoking the MIR limit, Gurvitch argued that no scattering event can take place before a charge carrier has moved a distance comparable to the lattice spacing. This then introduces the concept of a minimum time $\tau_{\text{min}}$ into the probability distribution,

$$\mathrm{d}P' = \begin{cases} 0, & t < \tau_{\text{min}} \\ \dfrac{1}{\tau} e^{-(t-\tau_{\text{min}})/\tau} \mathrm{d}t, & t \geq \tau_{\text{min}} \end{cases} \tag{5}$$

The inclusion of this 'forbidden zone' for scattering creates an additional term in the (Drude) expression for drift velocity (and hence conductivity) proportional to $\tau_{\text{min}}$ and therefore consistent with the MIR limit. Upon inverting this expression, the PRF for resistivity saturation is naturally obtained. Unfortunately, such a purely classical approach (5) does not lend itself to extension into the frequency domain, especially for frequencies smaller than $1/\tau_{min}$ [T. W. Silk and A. J. Schofield, private communication]. A more physically realistic scenario is one in which scattering at arbitrarily small times is allowed but where the *mean time* has an offset equivalent to $\tau_{min}$, i.e.

$$dP = \frac{1}{\tau + \tau_{min}} e^{-t/(\tau+\tau_{min})} \mathrm{d}t \tag{6}$$

Hence the mean scattering time has a minimum imposed by the presence of the lattice. This particular distribution clearly leads to the PRF at zero frequency, but in contrast with the staggered distribution of Gurvitch (1981), it can also be extended into the frequency realm. Its implementation is actually rather trivial; we simply insert $\tau_{min}$ ($\Gamma_{max}$) into the extended Drude expression for $\sigma(T,\omega)$ [Allen and Mikkelsen 1977], i.e.

$$\sigma(T,\omega) = \frac{\varepsilon_0 \Omega_p^2}{(\Gamma_{eff}(T,\omega) - i\omega[1+\lambda(T,\omega)])} \tag{7}$$

where $\lambda(T,\omega)$ is the temperature- and frequency-dependent coupling constant, obtained from $\Gamma_{eff}(T,\omega)$ via a Kramers-Kronig analysis and $\Gamma_{eff}(T,\omega)$ is the *effective* scattering rate given by

$$\frac{1}{\Gamma_{eff}(T,\omega)} = \frac{1}{\Gamma_{ideal}(T,\omega)} + \frac{1}{\Gamma_{max}} \tag{8}$$

For a standard metal, $\Gamma_{ideal}(T,\omega) = \Gamma_0 + \Gamma_{e\text{-}ph}(T,\omega)$, with $\Gamma_{e\text{-}ph}(T,\omega)$ given by the standard theory [Allen 1975], whilst for metals dominated by e-e scattering, $\Gamma_{ideal}(T,\omega) = \Gamma_0 + \Gamma_{e\text{-}e}(T,\omega)$ and $\Gamma_{e\text{-}e}(T,\omega) \sim ((\pi T)^2 + \omega^2)$. We stress once again that there is no second conducting channel here; it is the scattering rates that effectively add in parallel here, *not* the conductivities.

The effect of invoking a saturating rate ceiling on the optical response is illustrated in figure 2. Figure 2a shows $\sigma_{ideal}(T,\omega)$, the ideal conductivity spectra (i.e. with $\Gamma(T,\omega) = \Gamma_{ideal}(T,\omega) = \Gamma_0 + \Gamma_{e\text{-}ph}(T,\omega)$) for three representative temperatures; one on the coherent side ($T$ = 100K, $\rho_{ideal}(T) < \rho_{sat}$), one at the coherent/incoherent boundary ($T$ = 240K, $\rho_{ideal}(T) \sim \rho_{sat}$) and one for $\rho_{ideal}(T) >> \rho_{sat}$ ($T$ = 2400K). It is immediately clear from this figure that the high-$T$ $\sigma_{ideal}$ curve does not represent the experimental situation. The corresponding SRS spectra $\sigma_{SRS}(T,\omega)$ are given in figure 2b. The parameters used in these simulations are given in the figure caption. (Similar curves are also obtained using $\Gamma_{ideal}(T,\omega) = \Gamma_0 + \Gamma_{e\text{-}e}(T,\omega)$ though are not shown here). As we might expect, the influence of $\Gamma_{max}$ on $\sigma_{SRS}(T,\omega)$ grows as $T$ is increased. Note that since $\Gamma$ is of order $W$ at high $T$ and $\omega$, an ever increasing fraction of the optical spectra is shifted to $\omega > W$, thus violating the optical $f$-sum rule [Wooten 1972].

$$\int_0^W \sigma_1(\omega)d\omega = \frac{\pi\varepsilon_0\Omega_p^2}{2} \tag{9}$$

This violation is obviously more severe for $\sigma_{ideal}(T,\omega)$ than for $\sigma_{SRS}(T,\omega)$ since $\Gamma_{ideal}(T,\omega) > \Gamma(T,\omega)$ at all $T$ and $\omega$. In bad metals, the $f$-sum rule is clearly not conserved below $\omega = W$. The coherent/incoherent crossover around $T = T_{crit}$ is heralded by a flattening of the Drude response at low frequencies, followed by the development of a dip whose position shifts to higher $\omega$ with increasing $T$. The missing low-frequency spectral weight is transferred to energies beyond the bandwidth. For the HF metals (and presumably other saturating metals too), $W$ is the highest energy scale in the problem and consequently, spectral weight redistribution is confined to frequencies below $\omega = W$ (provided $T < W$) leaving the $f$-sum rule intact. Neither of these evolving spectra can be reproduced by our simple model and clearly one would need a much more sophisticated treatment to account for the redistribution of spectral weight in both cases in a self-consistent fashion. Nevertheless, several features of the plots derived from the SRS analysis are consistent with the experimental data, including the overlapping of spectra taken at different temperatures beyond $\omega \sim W/3$.

Taking our lead from HF metals, we summize that the Drude response in saturating metals must transform itself, if only asymptotically, into a flat spectrum of height $\sigma_{MIR} = \varepsilon_0\Omega_p^2/\Gamma_{max}$ extending all the way out to $\hbar\omega \sim W$ [Calandra and Gunnarsson 2002]. For such a case, we can invoke the $f$-sum rule (9) again to find an alternative expression for $\Gamma_{max}$, namely

$$\hbar\Gamma_{max} \approx \frac{2W}{\pi} \tag{10}$$

This particular definition of the MIR limit is rather general, is independent of dimensionality and does not involve the Fermi velocity $v_F$ whose absolute value in general is difficult to quantify. It is also consistent with the values found in high-$T_c$ cuprates [van der Marel *et al.* 2003] and those listed in table 2 for HF metals.

One implication of the above analysis is that $\rho(T)$ is going to deviate from $\rho_{ideal}(T)$ at temperatures well below saturation. This low-$T$ deviation from $\rho_{ideal}(T)$ implies that the charge carriers are somehow 'aware' of the lattice-imposed resistivity ceiling *even in the fully coherent region* where $\ell > 10a$, i.e. where Boltzmann theory still is a valid framework. We can understand this by noting that while there is a distribution of scattering times, with mean $\tau$, the presence of $\tau_{min}$ will be felt at all finite temperatures *irrespective of the value of $\tau$ itself*. Of course, for good metals such as copper, $\rho_{\rm ideal}(T)$ is always much smaller than $\rho_{\rm sat}$ so any departure from $\rho_{\rm ideal}(T)$ will be negligible at relevant temperatures.

In bad metals where scattering is much stronger, the ceiling in $\Gamma(T,\omega)$ should influence the transport data over a very wide temperature (and frequency) range, thus masking the *intrinsic* $T$-($\omega$-)dependence of the scattering rate, i.e. $\Gamma_{\rm ideal}(T, \omega)$. In high-$T_c$ cuprates, for example, it is widely believed that the observation of a $T$-linear resistivity, up to $T$ = 1000K in some cases [Gurvitch and Fiory 1987, Takagi *et al.* 1992], is evidence of some non-FL $T$-linear scattering rate induced by proximity to a quantum critical point [Varma *et al.* 1989]. In this paper, we have demonstrated how this association of a $T$-linear resistivity with a $T$-linear scattering rate is no longer appropriate, not only for resistivity values beyond the MIR limit, but also for those in its vicinity. Intriguingly, a recent phenomenological treatment for the cuprates that uses equation (8) with a $\Gamma_{\rm ideal}(T)$ that is *quadratic in temperature* (and anisotropic within the *ab*-plane) has been shown to generate both a $T$-linear resistivity and an inverse Hall angle $\cot\theta_H(T) \sim A + BT^2$ in optimally doped cuprates [Hussey 2003]. Since this issue is fundamental to unravelling the physics of high-$T_c$ cuprates, it will be explored in more detail in a forthcoming publication.

**5. CONCLUSIONS**

In this paper, we have tried to demystify bad metallic behaviour by adopting a broad empirical approach. In the process, we have uncovered a means of classifying saturating and non-saturating metals in terms of their optical response. We have found surprising parallels between bad metals and saturating metals which suggest that the original arguments of Mott, Ioffe and Regel may in fact be wholly inclusive. Bad metallic behaviour is associated largely with the absence of a zero-frequency collective mode. The development of the low-frequency dip above $T_{\rm crit}$ indicates a clear breakdown of the simple Drude picture. It implies that the continuation of the positive $\rho(T)$ slope into the incoherent regime has nothing to do with an unbounded escalation of the scattering rate, or in terms of the self-energy $\Sigma$, an ever-broadening spectral function. This conclusion is supported by the observation of saturation of $\Gamma(T,\omega)$ at high frequencies at a value $\Gamma_{\rm max} \sim W$. Thus, a description of bad metallic behaviour in terms of exotic scattering is no longer deemed appropriate. The definition of true metallic behaviour has to be confined, in semiclassical language, to charge carriers which move an *average* distance greater than the lattice spacing between collisions, or in quantum-mechanical language, to self-energies smaller than the bandwidth. It would be instructive in this regard to learn whether spectroscopic probes such as photoemission can detect such saturation in the quasiparticle self-energies of bad metals in future studies.

The crucial difference between saturating and non-saturating metals appears to lie in the preservation (or loss thereof) of spectral weight within a cut-off associated with the conduction bandwidth. In saturating metals, as exemplified by the HF metals, the total spectral weight is preserved and at temperatures in the saturation limit, $\sigma(T,\omega)$ is characterized by a flat, featureless spectrum extending all the way out to $\hbar\omega \sim W$. In bad metals, the optical sum rule is violated beyond $T_{crit}$ with the missing low-frequency spectral weight becoming redistributed to energies beyond $\hbar\omega = W$, as described in DMFT. Evidence suggests however that this coherent/incoherent crossover in the optical response is governed by the same MIR criterion for dc conductivity as found in saturating metals.

We have tried to adapt Gurvitch's original ideas, linking resistivity saturation to saturation of the scattering rate at the MIR limit, to the frequency domain. Whilst such SRS analysis explains some features of the data, e.g. the weak $T$-dependence of the spectra beyond $\omega \sim W/3$, the modelling clearly requires greater sophistication (e.g. to incorporate the spectral weight redistribution) before it can be applied more meaningfully to the full energy range. We do believe however that scattering rate analysis of the type presented here must now be incorporated in any detailed theory of the optical response of strongly scattering metals. As mentioned previously, one important implication of the SRS analysis is that $\rho(T)$ will deviate from $\rho_{ideal}(T)$ at temperatures well below $T_{crit}$. It is therefore very likely that in many bad metals, the presence of $\Gamma_{max}$ will influence the transport data over a very wide temperature (and frequency) scale. The impact of this on our current understanding of the intrinsic behaviour of $\Gamma(T,\omega)$ in bad metals in general, and cuprates in particular, is potentially far-reaching.

**ACKNOWLEDGEMENTS**

The authors would like to acknowledge fruitful discussions with N. Bontemps, A. J. Schofield, T. W. Silk, T. Timusk, D. van der Marel, J. A. Wilson, Y. Yanase and J. Zaanen. We also thank J. Alexander for preparing figure 2 and J. A. Wilson for his critical reading of the manuscript.

**APPENDIX**

**Resistivity Saturation in Carbon Fullerenes**

In the alkali-doped fullerenes, the issue of resistivity saturation has received significant attention. On the theoretical side, it has been suggested that the fullerenes are rather a unique case, in which strong coupling to intramolecular vibrations leads to a lack of resistivity saturation with mean-free-paths well below the MIR limit [Gunnarsson and Han 2000]. Since electron-phonon scattering is believed to be the dominant scattering mechanism here, this result would appear to contradict our proposed phenomenology. The experimental situation, however, remains controversial with several experimental issues still awaiting resolution. Firstly, whilst some reports show resistivity data extending to values up to 6 mΩcm without saturation [Hebard *et al*. 1993], others show the onset of resistivity saturation (and PRF scaling) but with a small $\ell_{min}$ ~ 1Å << *a* [Hou *et al*. 1995]. Secondly, it has been notoriously difficult to determine the absolute value of the resistivity in these compounds. Indeed, resistivity values reported in the literature [Gunnarsson 1997] differ by one order of magnitude. Thirdly, knowledge of the intrinsic *T*-dependence of $\rho(T)$ in fullerenes is complicated by thermal expansion effects and a *T*-dependent density of states. Finally, optical data by Degiorgi *et al*. (1994) show a small but sharp Drude peak appearing below 100 $cm^1$ (at low *T*), suggesting that the high resistivity values may be due to a low concentration of mobile carriers. If this is the case, the minimum conductivity (maximum resistivity) would be much lower (higher) than originally thought. Contrasting data by Iwasa and Kaneyasu (1995) suggest a much larger and broader Drude response with an anomalously high $\Gamma$ ~ 1 eV. It is noteworthy however that the Iwasa measurements stop at 100 $cm^{-1}$, precisely the point at which Degiorgi *et al*. observe the sharp (Drude) upturn. Further measurements of the optical response of fullerenes, especially to higher temperatures, and independent estimates of the carrier density, e.g. via complementary Hall effect measurements, would be very welcome. We therefore reserve to make a definitive statement about the alkali-doped fullerenes to a later date, but acknowledge that their reported behaviour is not at present consistent with our classification scheme.

## FIGURE CAPTIONS

Figure 1: Temperature-dependent in-plane optical conductivity spectra $\sigma_{ab}(T,\omega)$ for $La_{1.9}Sr_{0.1}CuO_4$ at specified temperatures. Inset: In-plane resistivity data for $La_{1.9}Sr_{0.1}CuO_4$ up to 1000K [Taken from Takenaka *et al*. (2003)].

Figure 2(a): Simulation of the ideal optical conductivity $\sigma_{\rm ideal}(T,\omega)$ at 100K (red), 240 K (blue) and 2400K (green) assuming $\Gamma_{\rm ideal}(T,\omega) = \Gamma_0 + \Gamma_{\rm e\text{-}ph}(T,\omega)$, with $\Gamma_0$ = 300K and $\Gamma_{\rm e\text{-}ph}(T,\omega)$ given by the standard expression [Allen 1965]. In this example, $\theta_D$ = 400K and the electron-phonon spectral function $\alpha^2F(\nu) = 5 \times 10^{-9}\ \nu^4$. 2(b): Simulation of $\sigma_{\rm SRS}(T,\omega)$ using equation (8) with $\Gamma_{\rm max}$ = 3000K. $\Omega_p$ is assumed to be $\omega$-independent and has been given an arbitrarily small value (= 4000K) in order to exaggerate the effects of $\Gamma_{\rm max}$ on $\sigma_{\rm SRS}(T,\omega)$.

## TABLE CAPTIONS

Table 1: Selection of metallic elements and alloys for which the parallel resistor formula (eqn (1)) has been successfully applied, together with corresponding values of $\rho_{\rm sat}$ and $\ell_{\rm min}$ (in units of *a*). As shown by Gurvitch (1981), the range of $\rho_{\rm sat}$ values can be accounted for largely by variations in the carrier density *n*.

Table 2: Resistivity and optical conductivity parameters derived from experiments in three heavy-fermion metals, $UPt_3$, $UPd_2Al_3$ and $CeAl_3$. In cases where the resistivity is anisotropic, an average value for the three crystal axes is shown. The *W* values were estimated from the mid-point in the tail-off of the high-frequency intra-band conductivity. Note that these values may be subject to large uncertainty due to emerging contributions from inter-band transitions.

Table 3: Selection of bad metals with corresponding parameters obtained from resistivity and optical conductivity measurements, as explained in the text. Note that for $La_{2-x}Sr_xCuO_4$, $\sigma_{\rm crit}(0)$ remains largely independent of *x*. The asterix symbols refer to predicted values.

| Compound | $\rho_{sat}$ (μΩcm) | $\ell_{min}$ ($a$) | Reference |
|---|---|---|---|
| <u>Elements</u> | | | |
| La | 175 | ~ 1 | [Sundqvist 1992] |
| Ti | 200 | ~ 1 | [Lin *et al*. 1993] |
| <u>A15 compounds</u> | | | |
| $Nb_3Sn$ | 135 -150 | ~ 1 | [Fisk *et al*. 1976]<br>[Wiesmann *et al*. 1977] |
| $Nb_3Ge$ | 135 | ~ 1 | [Wiesmann *et al*. 1977] |
| $Nb_3Pt$ | 105 | ~ 1 | [Caton and Viswanathan 1982] |
| $V_3Si$ | 135 | ~ 1 | [Caton and Viswanathan 1982] |
| $Mo_3Ge$ | 120 | ~ 1 | [Gurvitch *et al*. 1978] |
| <u>Alloys and Intermetallics</u> | | | |
| $Ti_{1-x}Al_x$ | 190 – 330 | 1 – 1.5 | [Mooij 1973]<br>[Lin *et al*. 1993] |
| $In_5B_3$ | 140 | 1 – 1.5 | [Mori *et al.* 1981] |
| $Fe_{80-x}Ni_xCr_{20}$ | 180 – 250 | 1 – 1.5 | [Nath and Majumbar 1996] |

Table 1

| Compound | $\rho_{sat}$ (μΩcm) | $\Omega_p$ (ev) | $\Gamma_{max}$ (cm$^{-1}$) | $W$ (cm$^{-1}$) | $\sigma_{min}$ (Ωcm$^{-1}$) | $a$ (Å) | Reference |
|---|---|---|---|---|---|---|---|
| $UPt_3$ | 200 | 2.6 | 1500 | 2500 | 5000 | 5 | [de Visser *et al*. 1984]<br>[Sulewski *et al*. 1988]<br>[Dressel *et al*. 2002] |
| $UPd_2Al_3$ | 150 | 3.2 | 1600 | 1500 | 6000 | 4 | [Geibel *et al*. 1991]<br>[Dressel *et al*. 2002] |
| $CeAl_3$ | 60 | 3.5 | 1050 | 1800 | 13000 | 4 | [Awashti *et al*. 1993] |

Table 2

| Compound | $\rho_{met}$ ($\mu\Omega$cm) | $T_{crit}$ (K) | $\sigma_{crit}(0)$ $(\Omega cm)^{-1}$ | $\sigma_{MIR}$ $(\Omega cm)^{-1}$ | Reference |
|---|---|---|---|---|---|
| $La_{1.94}Sr_{0.06}CuO_4$ | 5500 | 200 | 800 | 1400 | [Takagi *et al*. 1992] [Takenaka *et al*. 2002a, 2003] |
| $La_{1.92}Sr_{0.08}CuO_4$ | 4000 | 295 | 1000 | 1450 | [Takagi *et al*. 1992] [Takenaka *et al*. 2002a, 2003] |
| $La_{1.9}Sr_{0.1}CuO_4$ | 4000 | 400 | 1000 | 1500 | [Takagi *et al*. 1992] [Takenaka *et al*. 2002a, 2003] |
| $YBa_2Cu_4O_8$ | 1100 | 300-450* | - | 1500* | [Bucher *et al*. 1993] [Hussey *et al*. 1997] |
| $La_{0.6}Sr_{0.4}MnO_3$ | 2000 | 325 | 1500 | 1000 | [Takenaka *et al*. 2002b] |
| $La_{0.825}Sr_{0.175}MnO_3$ | 10000 | 175 | 1500 | 1000 | [Takenaka *et al*. 1999] |
| $Li_2VO_4$ | 2500 | 250* | - | 1500* | [Urano *et al*. 2000] |
| $Sr_2RuO_4$ | 1300 | 400* | - | 4000* | [Tyler *et al*. 1998] |
| $SrRuO_3$ | 300 | 145 | 9000 | 5000 | [Allen *et al*. 1996] [Kostic *et al*. 1998] |

Table 3

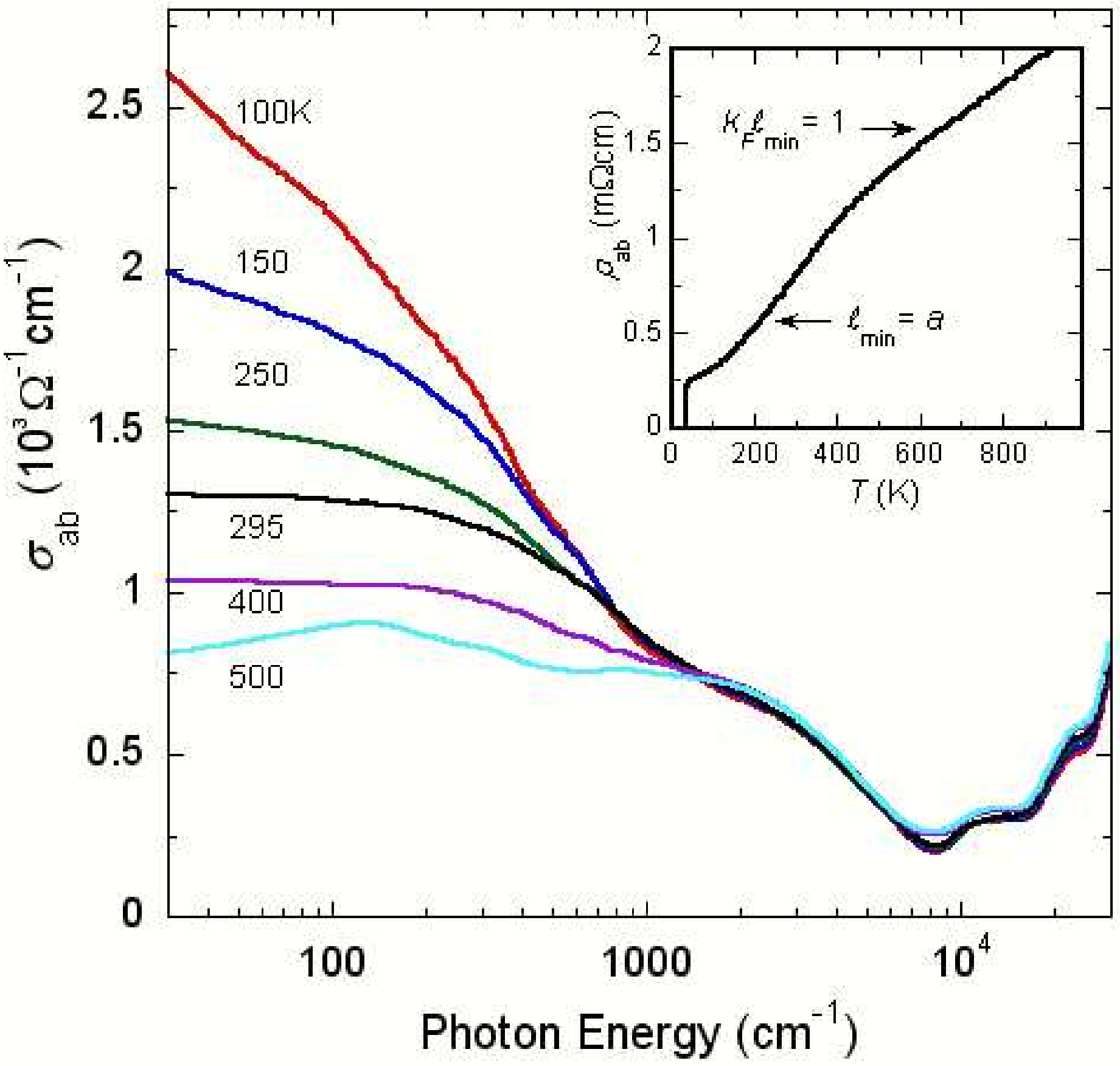

Figure 1

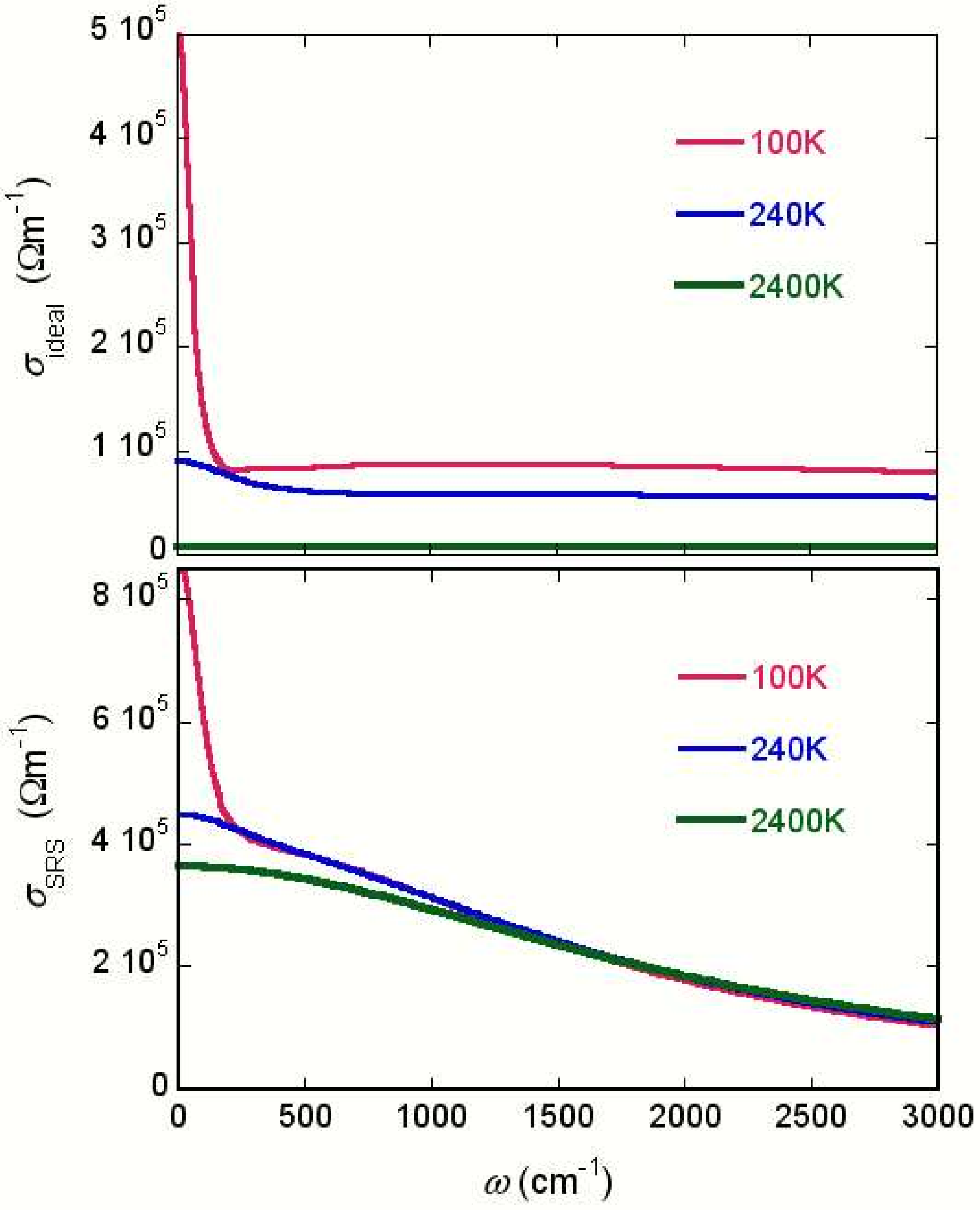

Figure 2